\documentclass[aps,pre,preprint]{revtex4}
\usepackage{latexsym}
\usepackage{natbib}
\usepackage{amsmath}
\usepackage{amsfonts}
\usepackage{amssymb}
\usepackage[T1]{fontenc}
\usepackage[dvips]{color,graphicx}   
\usepackage{verbatim}   
\usepackage{t1enc}
\usepackage{anysize}
\usepackage{tocbibind}
\usepackage{subfig}
\usepackage{graphicx}
\usepackage[prrcent]{overpic} 
\usepackage{float}  
\usepackage{appendix} 
\usepackage{booktabs} 
\usepackage{natbib}
\usepackage{multirow}
\usepackage{array}
\usepackage{color}
\newcommand{\overbar}[1]{\mkern 1.5mu\overline{\mkern-1.5mu#1\mkern-1.5mu}\mkern 1.5mu}
\DeclareGraphicsExtensions{.pdf,.png,.jpg}

\begin{document}
\widetext
\title{Recovery of Interdependent Networks}

\author{M. A. Di Muro}
\affiliation{Instituto de Investigaciones F\'isicas de Mar del Plata
  (IFIMAR)-Departamento de F\'isica, Facultad de Ciencias Exactas y
  Naturales, Universidad Nacional de Mar del Plata-CONICET, Funes
  3350, (7600) Mar del Plata, Argentina.} 
 \author{C. E. La Rocca}
\affiliation{Instituto de Investigaciones F\'isicas de Mar del Plata
  (IFIMAR)-Departamento de F\'isica, Facultad de Ciencias Exactas y
  Naturales, Universidad Nacional de Mar del Plata-CONICET, Funes
  3350, (7600) Mar del Plata, Argentina.}  \author{H. E. Stanley}
\affiliation{Center for Polymer Studies, Boston University, Boston,
  Massachusetts 02215, USA} \author{S. Havlin} \affiliation{Minerva
  Center, Bar Ilan University, Ramat Gan, Israel.}
\author{L. A. Braunstein} \affiliation{Instituto de Investigaciones
  F\'isicas de Mar del Plata (IFIMAR)-Departamento de F\'isica,
  Facultad de Ciencias Exactas y Naturales, Universidad Nacional de
  Mar del Plata-CONICET, Funes 3350, (7600) Mar del Plata, Argentina.}
\affiliation{Center for Polymer Studies, Boston University, Boston,
  Massachusetts 02215, USA}
\date{\today} 

\begin{abstract}
Recent network research has focused on the cascading failures in a
system of interdependent networks and the necessary preconditions for
system collapse. An important question that has not been addressed is
how to repair a failing system before it suffers total breakdown. Here
we introduce a recovery strategy for nodes and develop an analytic and
numerical framework for studying the concurrent failure and recovery
of a system of interdependent networks based on an efficient and
practically reasonable strategy. Our strategy consists of repairing a
fraction of failed nodes, with probability of recovery $\gamma$, that
are neighbors of the largest connected component of each constituent
network. We find that, for a given initial failure of a fraction
$1-p$ of nodes, there is a critical probability of recovery above
which the cascade is halted and the system fully restores to its
initial state and below which the system abruptly collapses. As a
consequence we find in the plane $\gamma-p$ of the phase diagram three
distinct phases. A phase in which the system never collapses without
being restored, another phase in which the recovery strategy avoids
the breakdown, and a phase in which even the repairing process cannot
prevent system collapse.
\end{abstract}

\flushbottom
\maketitle

\thispagestyle{empty}

\section*{Introduction}

In recent years researchers have attempted to understand the
topological structure and self-organization of complex systems.  The
field of complex networks, which characterizes components of a complex
system as nodes and their interactions as links, has emerged as a
natural outgrowth of this quest. Studies of the Internet, human and
animal societies, climate systems, physiological systems,
transportation systems, biochemical reactions, and food webs in
ecosystems are only few examples of systems that are better understood
using complex network
theory~\cite{Alb_00,Bar_01,Bas_01,Coh_01,Pas_01b,Dor_03,Bra_03,New_08,Cal_07,Alo_07,Goz_08,Don_09,New_10,Coh_10,Liu_11,Sch_11,gli2010,DaqingPNAS,Cai}.
However, it was recently demonstrated that many complex systems cannot
be described adequately as single isolated networks but should be
represented as interdependent networks, which are characterized by
connectivity links within each network and dependency links between
networks \cite{Bul_01,Vespi2010}. Technological infrastructures
provide the most obvious examples.  Electrical, gas, and water
networks rely on telecommunications networks for their control
systems. Water systems are used to cool generators in an electrical
system.  Nearly every infrastructure network depends on the power grid
to function. Such macro systems are much more complex and vulnerable
compared to isolated networks.  For interdependent networks the
distinction between \emph{internal} connectivity links within each
network and \emph{interdependent} links between the networks
represents new challenges and the interest and research in these
multiple coupled systems has recently rapidly expanded.  In September
2003 a tree fell on a transmission line in Switzerland and triggered a
cascade of failures that left 53 million people in the dark, most of
them in Italy. This additional massive blackout to the growing list of
global large-scale catastrophic events has motivated the study of
robustness and cascading failures in interdependent networks. Using
percolation theory, Buldyrev \emph{et al.}~\cite{Bul_01} developed a
framework for studying interdependent networks and found that the
coupled system behaves very different from a single isolated network
and is significantly more fragile.  In contrast to the percolation of
single networks where the transition is in general continuous, an
abrupt first-order percolation transition was found in interdependent
networks where near the critical point a tiny fraction of node
failures can cause cascading failure and system collapse.

It was also found \cite{Par_02} that reducing interdependencies between
networks below a critical value yields a continuous percolation
transition. Very recently, Gao {\it et al.}~\cite{gao2011}, Schneider
{\it et al.}~\cite{Sch_03} and Valdez {\it et al.}~\cite{Val_jpa} showed
that backing up high-degree interdependent nodes enhances the robustness
of a coupled system. It was also found that networks with assortative
dependency (i.e., nodes with similar degrees in both networks tend to be
dependent) are more robust than networks with random dependency
\cite{buldy_corr,Roni_corr,Val_pre}. Previous resilience studies have
focused on failure propagation and the breakdown of systems of coupled
networks. Much work has been devoted to the design of control and
mitigation strategies \cite{Par_02,buldy_corr,Roni_corr,Val_pre} to
avoid catastrophic events and to heal failures as they occur. In order
to reduce overload failures in power systems, some proposed control
strategies consist of simply strengthening the capacity or reducing the
load of groups of nodes. Mitigation can also be achieved by
``islanding'' nodes, i.e., separating certain clusters from the main
power grid and powering them with independent alternative sources as
solar or wind power~\cite{Pat_10}. Nevertheless, in real-world scenarios
nodes can be repaired or recovered. Complex networks with heterogeneous
distribution of loads may undergo a global cascade of overload failures
when highly loaded nodes or edges are removed due to attacks or
failures. Since a small attack or failure has the potential to trigger a
global cascade, a fundamental question of much interest is regarding the
possible strategies of defense to prevent the cascade from propagating
through the entire network.  Motter \cite{motter_prl} introduced a
strategy of defense to prevent a global cascade of overload failures in
isolated heterogeneous networks using a selective removal of nodes and
edges right after the initial attack or failure. This intentional
removal of network nodes and edges drastically reduces the size of the
cascade. Majdandzic {\it et al.}  \cite{antonio} studied a failure
recovery model in isolated networks where the failures are due to lack
of support within the networks. In \cite{antonio} after an inactive
period of time a significant part of the damaged network is capable, due
to internal fluctuations, of spontaneously becoming active
again. However, repairing interdependent networks that experience a
cascade of failures is a possibility that has not yet been taken into
consideration.

In this work, we develop a model for the competition between the
cascading failures and the restoration strategy that repairs failed
nodes in the boundary of the functional network and reconnects them to
it (see Fig.~\ref{shlomo}). The reasoning behind this repairing strategy
is based on the fact that (a) in many real systems it is easier to
repair boundary nodes (for example, in a transportation system one needs
to bring equipment to the damaged site and it is easier to bring it near
using the existing transportation system) and (b) fixing a node that is
not in the boundary will cause the node to fail in the next step since
it is not connected to the giant component and thus such a repair will
be a wasted effort. In order to determine the recovery probability
necessary to protect a system from collapse, we develop a theoretical
model that is solved using random percolation theory. We present
numerical solutions for the evolution of the theoretical process as well
as for the steady states and compare them with simulations. We find that
there is a critical probability $\gamma_c$ that depends on $p$ that
separates a regime of full system fragmentation from a regime of
complete system restoration.

\section*{Results}

\subsection*{Model}
\smallskip
For the sake of simplicity and without loss of generality, we
consider two interdependent networks $A$ and $B$.

\subsubsection*{Stochastic Model}
Both networks have the same number of nodes $N$.  Within each network
the nodes are randomly connected through connectivity links with a
degree distribution $P^i(k)$, where $i = A$ or $B$. Pairs of nodes
across the two networks are randomly connected one-to-one via
bidirectional interdependent links as in Buldyrev {\it et
  al.}~\cite{Bul_01}.

We assume that at the initial stage a fraction $1-p$ of nodes in network
$A$ fail. The failure spreads in network $A$ through connectivity links
and all the nodes that do not belong to the functional giant component
(GC) of network $A$ fail and it is assumed that they become
dysfunctional. The failed nodes in network $A$ no longer support their
corresponding nodes in network $B$ through their interdependent links,
and those nodes in network $B$ that were dependent on the failed nodes
in $A$ also fail. If the fraction of the initially-failed nodes in $A$
is above $1-p_c$, where $p_c$ is the critical threshold, and there is no
repair strategy, a catastrophic cascade of failures occurs and the
system abruptly collapses.  Our model assumes a process of recovery that
is immediately applied at the first step of the cascade of failures with
the objective of avoiding or delaying the collapse of the system. In
this process certain failed nodes are recovered according to the
following rules:

(i) If a failed node in one network is at a distance $\ell=1$ from its
GC (we denote the collection of nodes at distance $\ell=1$ from the GC
as the boundary of the GC) and has an interdependent link with a
failed node in the other network that is also at a distance $\ell=1$
from its corresponding GC, this pair of nodes belongs to the mutual
boundary and the two are repaired with a probability $\gamma$.

\begin{figure}[h]
\centering
\includegraphics[scale=0.28]{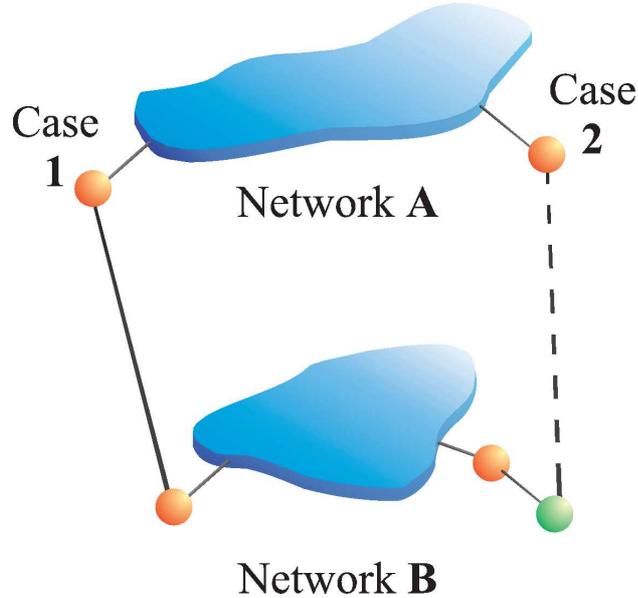}
\caption{(Color online) Schematic rules of the failure-recovery
  strategy. The GCs of networks $A$ and $B$ are shown (blue). In orange
  we mark boundary nodes at a distance $\ell=1$ from their respectively
  GCs and in green a node with a distance $\ell =2$ from the GC in
  B. Case 1: Two interconnected failed nodes at a distance $\ell=1$ from
  their respectively GCs are repaired with probability $\gamma$. Case 2:
  If at least one of the two interconnected failed nodes is at a
  distance $\ell>1$ from its GC, we do not recover these nodes. Note
  that this type of recovery is practical and realistic, since in real
  infrastructure it is usually more convenient to repair boundary nodes
  which are next to the functional infrastructure GC.}
\label{shlomo}
\end{figure}

(ii) If the interdependent node in the other network does not belong
to the boundary, none of them is repaired. Figure~\ref{shlomo}
sketches the recovery strategy.

It is important to clarify that when a node of the boundary is
restored, not only all its connections with the GC are reactivated,
but also its connectivity links with other restored nodes from the
same network are recovered (if they were connected originally).

We denote by $n= 0, 1, ....$ the time steps of the cascading
process. In the simulations at $n=0$ a fraction $1-p$ of nodes fails
in network $A$. From the fraction $p$ of nodes that survive, only
those within the GC are regarded as functional while the others are
dysfunctional and considered as failed nodes. After the initial
failure the damage in $A$ propagates to network $B$ through the
interdependent links, as the conventional process of cascading
failures introduced in Ref~\cite{Bul_01}, but before spreading the
failures back to network $A$ we restore the interdependent nodes that
belong to the mutual boundary of both networks $A$ and $B$ with a
probability $\gamma$. The rules of the model for any stage $n$ are
given by:

\begin{itemize}
  \item Stage $n$ in $A$
    \begin{enumerate}
         \item Functional nodes fail if they lose support from their
           counterpart nodes in $B$ at stage $n-1$.
         \item From the survivors, those nodes that belong to the GC
           of $A$ remain functional while the others fail.

    \end{enumerate}

  \item Stage $n$ in $B$:

    \begin{enumerate}

         \item Functional nodes become dysfunctional if they lose
           support from network $A$ due to the cascade of failures at
           stage $n$.
         \item The remaining nodes fail if they do not belong to the
           GC of $B$.
         \item Interdependent nodes in the mutual boundary of the GCs
           of networks $A$ and $B$ are restored with probability
           $\gamma$. All their connections with the respective GC are
           reactivated and also the links between restored boundary
           nodes, if they were connected before the failure.
    \end{enumerate}

  \end{itemize}
  
This procedure is repeated until a steady state is reached, which
depends upon $\gamma$ and $p$. In this state there are no finite
clusters in any network and the fraction of nodes that belongs to the
GC, $P^i_\infty$, $i=A,B$ in both networks, is the same because any
node in each network is supported through interdependent links by
the other node in the other GC. 

\subsubsection*{Theoretical Approach}

In order to solve our theoretical model, we use the generating
function formalism \cite{New_05,Bra_01} extended to interdependent
networks \cite{Val_jpa,Val_pre,Bul_01,zhou,gao2011,roni_PNAS_2013},
which is based on two generating functions in which $G^i_0(y)~=~\sum_k
P^i(k) y^k$ is the generating function of the degree distribution,
$G^i_1(y)=\sum_k k P^i(k) \; y^{k-1}/\langle k_i \rangle$ is the
generating function of the excess degree distribution, and $\langle
k_i \rangle$ is the average degree of the network, with $i=A,B$. Using
this formalism, we denote by $g_A[x]$ ($g_B[y]$) the order parameter
$P^A_\infty \equiv P^A_\infty[x] $ ($P^B_\infty \equiv
P^B_\infty[x]$~) evaluated at $x$ ($y$), then for the network $A$
\[P^A_\infty = g_A[x] = x \,(1-G_0^A [1- f^A_\infty]),\]
where $f^A_\infty \equiv f^A_\infty[x]$ satisfies the self-consistent
equation
\begin{equation}\label{Eq.finf}
f^A_\infty= x \,(1-G_1^A[1- f^A_\infty]),
\end{equation}
where $f^A_\infty$ is the probability that an infinite branch expands
the system in network $A$
\cite{roni_PNAS_2013,zhou,gao2011,Bra_01,Val_jpa,Val_pre}. The same
equations and definitions hold for network $B$ with
\[P^B_\infty= g_B[y] = y \,(1-G_0^B [1- f^B_\infty]),\]
and 
\[f^B_\infty= y \,(1-G_1^B[1- f^B_\infty])\ , \]
where also $f^B_\infty \equiv f^B_\infty[y]$.

As our theory is based on node percolation where finite clusters are not
regarded as functional, dysfunctional nodes are failed nodes and nodes
that belong to finite clusters are also failed. We denote by $p^{A}_{n}$
($ p_{n}^{B}$) the effective fraction of nodes remaining in network $A$
($B$) after the cascade of failures and before repairing at step $n$. At
stage $n=0$ we have a fraction $1-p$ of nodes from network $A$ that fail
and therefore $p_{0}^{A}=p$ and $p_0^B= g_A[p]$ (for a detailed
description of the process see {\it Supplementary Information:
  Theory}). After the initial cascade that goes from network $A$ to
network $B$, the process of recovery begins. At stage $n$ the fraction
of nodes in the GC of networks $A$ and $B$ is given by
\begin{eqnarray}\label{pinfa}
P^A_{\infty,n}& =& g_A[p_{n}^{A}]\ ;\nonumber\\
P^B_{\infty,n}& = &g_B[p_{n}^{B}]\ .
\end{eqnarray}
\begin{figure}[ht]
\centering
\includegraphics[scale=0.9]{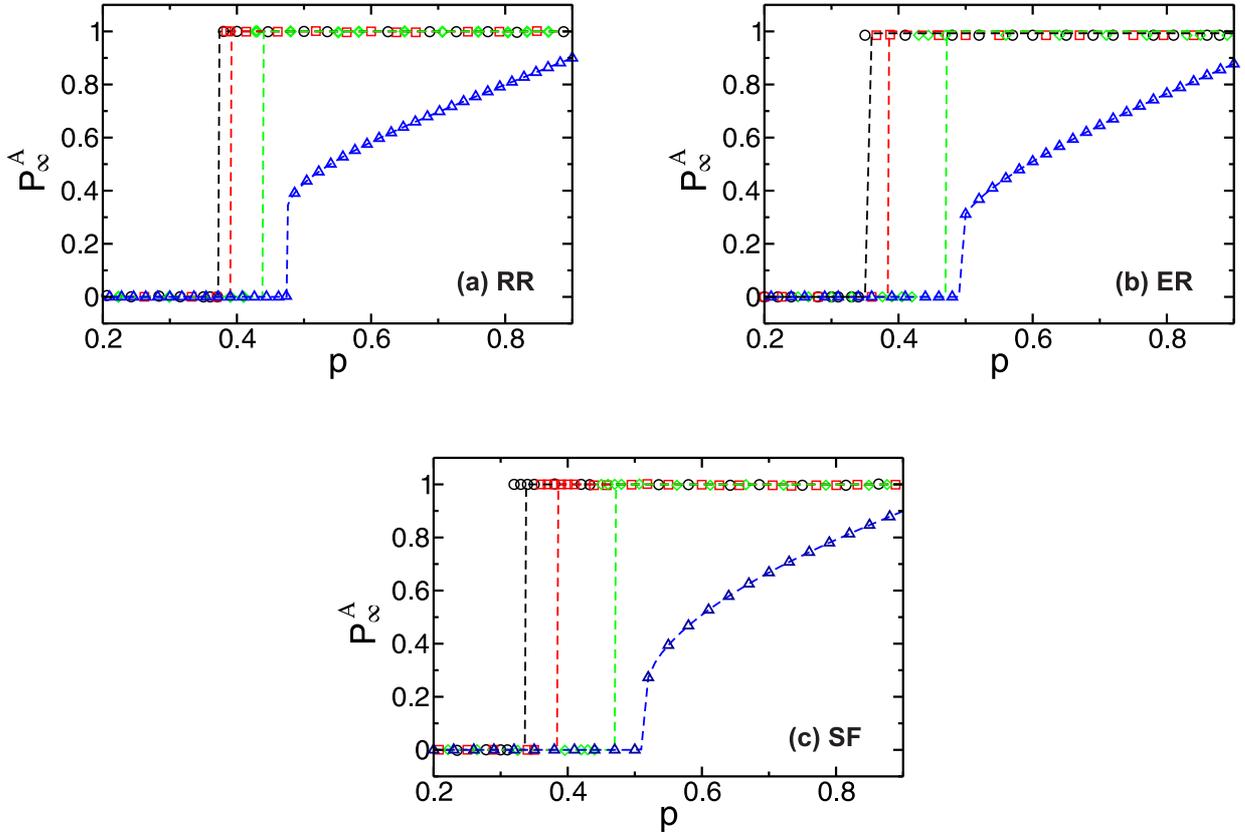}
\vspace{0.2cm}
\caption{(Color online). Fraction of nodes in the GC of network $A$,
  $P^A_{\infty}$, in the steady state of network $A$ as a function of
  $p$ for $N=10^6$ with $\gamma=0$ ($\bigtriangledown$), $\gamma=0.1$
  ($\Diamond$), $\gamma=0.5$ ($\Box$) and $\gamma=1$ ($\bigcirc$) for
  (a) RR networks with $z=5$, (b) ER networks with $\langle k \rangle
  =5$ and (c) SF networks with $\lambda=3.0$, with lower and maximal
  connectivity $3$ and $1000$ which correspond to $\langle k \rangle
  \approx 5.11$. We include as a reference the no recovery case,
  $\gamma=0$. The symbols correspond to the simulations and the dashed
  lines are the theoretical solutions of
  Eqs.~(\ref{pinfa}-\ref{Eq.pq}). Simulations have been averaged over
  $1000$ network realizations.}
\label{Pinf_sim_teo}
\end{figure}

The nodes that are repaired are those that belong to the mutual
boundary of both GCs. The fraction of nodes that are in the boundary
of each GC can be written as
\begin{eqnarray}\label{yyy01z}
F_{n}^{A}&=&(1-p^A_n)\;(1-G^A_0 [1- f^A_{\infty,n}])\ ;\nonumber\\
F_{n}^{B}&=&(1-p^B_n)\;(1-G^B_0 [1- f^B_{\infty,n}])\ ;
\end{eqnarray}
where the factor $(1-G^i_0 [1- f^i_{\infty,n}])$ is the probability
that a node is connected to the GC at stage $n$ and the factor
$(1-p^{i}_{n})$ is the fraction of nodes that fail, with $i=A,B$.
The mutual boundary, which is the fraction of nodes in the boundary of
network $B$ that are interconnected via dependency links to the nodes
in the boundary of network $A$ at stage $n$, can be written as
\begin{equation}
F_{n}^{AB}=F_{n}^{B}\frac{F_{n}^{A}}{1-g_A[p^{A}_{n}]}\ ,
\end{equation}
where $F_{n}^{A}/(1-g_A[p^{A}_{n}])$ is the conditional probability
that a node belongs to the boundary of the GC of network $A$ given
that it is interconnected via an interdependent link with a node that
belongs to the boundary of the GC of network $B$. 

Next we compute the fraction of nodes in the GCs, $\overbar
{P^i_{\infty,n}}$, after repairing at stage $n$ by adding a fraction
$\gamma$ of the mutual boundary to the values of Eqs.~(\ref{pinfa})
\begin{eqnarray}\label{Eq.pinfn}
\overbar{P^A_{\infty,n}} &=& g_A[p^A_n] + \gamma F_{n}^{AB}\ ,\nonumber\\
\overbar{P^B_{\infty,n}} &=& g_B[p^B_n] + \gamma F_{n}^{AB}\ ,
\end{eqnarray}
where the bar indicates the relative size of the order parameter of the
enlarged GCs due to the restoration process.

Finally, we compute the fraction of remaining nodes in each network
after the recovery process $q_n^A (q_n^B)$ by solving the pair of
transcendental equations
\begin{eqnarray}\label{Eq.q}
g_A[q_{n}^{A}]&=&\overbar{P^A_{\infty,n}}\ ,\nonumber\\
g_B[q_{n}^{B}]&=&\overbar{P^B_{\infty,n}}\ .
\end{eqnarray}
Then for any stage $n > 0$ the
fraction of nodes remaining after the cascade of failures and before
the repairing process in each network is given by
\begin{eqnarray}\label{Eq.pq}
p^A_n &=& q^A_{n-1} \frac{g_B[q^B_{n-1}]}{g_A[q^A_{n-1}]}\ ,\nonumber\\
p^B_n &=& q^B_{n-1} \frac{g_A[p^A_n]}{g_B[q^B_{n-1}]}\ .
\end{eqnarray}
The process is iterated until the steady state is reached, when there
are no more nodes belonging to the mutual boundary.  

In Fig.~\ref{Pinf_sim_teo} we show $P^A_\infty$ as a function of $p$
in the steady state for three different systems of interdependent
networks: two Random Regular (RR), two Erd\"{o}s R\'enyi (ER) and two
Scale Free (SF), characterized by a power law degree distribution with
exponent $\lambda=3$.  In these plots we show simulation results
(symbols) and theoretical results, presented below, (lines) for four
values of $\gamma$. The case of $\gamma=0$ is shown as a
reference. The details of the simulations are presented in {\it
  Section: Methods}. We can see that the critical threshold $p_c$
decreases when $\gamma$ increases, and thus the networks become more
robust. Note that in our restoring model a steady state is reached
when the system is either fully functional or fully collapsed and that
there is no intermediate state. In the {\it Supplementary Information:
  Analytical solutions for the fraction of nodes in the GCs} we show
that in our model the only solutions for the fraction of nodes in the
GC of both networks are either one or zero. This is in contrast to
other models of cascading failures in interdependent networks without
recovery
\cite{Bul_01,Par_02,Sch_11,Sch_03,Val_jpa,gli2010,zhou,roni_PNAS_2013,gao2011,Baxter}
where intermediate states exist.  From Fig.~\ref{Pinf_sim_teo} we can
see that the agreement between the theory and the simulations is very
good for all cases. We find that the case of coupled SF networks for
$\gamma > 0$ is the one that presents the largest deviation in $p_c$
compared with more homogeneous networks, such as the ER or the RR.
\begin{figure}[ht]
\centering
\includegraphics[scale=0.4]{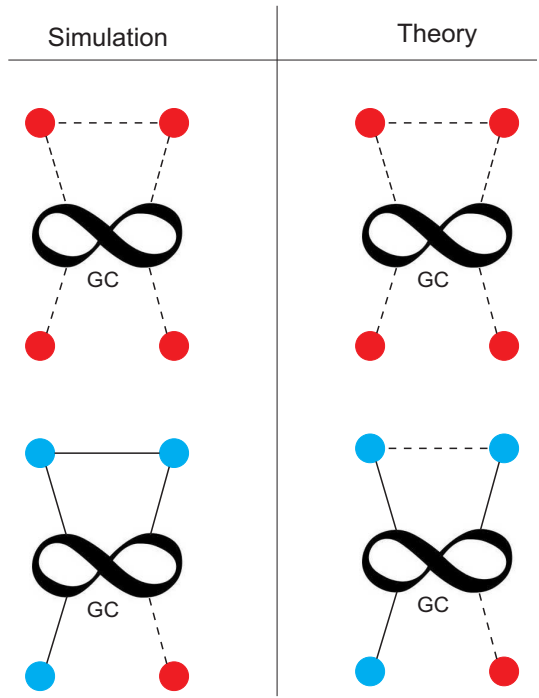}
\caption{(Color online) Schematic comparison of the rules of the
  recovery strategy between theory and simulations. In the left panel
  we show the rules for the simulations and in the right panel the
  rules for the theory. For simplicity we show only one of the coupled
  networks. The infinity symbol indicates the Giant Component (GC),
  red nodes are boundary failed nodes, while blue nodes are boundary
  recovered nodes. The dashed lines indicate inactive connections and
  the solid line reactivated connections. In the simulations a
  connection between two boundary recovered nodes is restored, which
  is not contemplated in the theory. In this case $\gamma=3/4$, and
  boundary nodes have only one connection to the GC for simplicity.}
\label{esquem}
\end{figure}
To explain this deviation note that in our analytic approach we map node
removal and repairing into random percolation. This means that in the
theory all nodes have the same probability of failure and recovery. The
repaired nodes are attached to the GC of their networks and cannot fail
in the next step of the cascade of failures. It can be shown that the
probability that a node belongs to the boundary increases with its
degree (see {\it Supplementary Information: Excess Degree of the
  Boundary}), and this effect is more pronounced as the heterogeneity of
the networks increases. In addition, recall that the simulation model
reactivates broken connections between boundary restored nodes. This
last feature of the model is illustrated in Fig.~\ref{esquem}. These
effects result in an increasing of the mean connectivity of the GC of
each network. Thus the process of border recovery in the simulation
generates a structure that is more resilient against failures than the
structure in the theoretical approach, and therefore the critical
thresholds of the theory are slightly higher from those of the
simulations. 

\begin{figure}[h]
\centering
\includegraphics[scale=0.8]{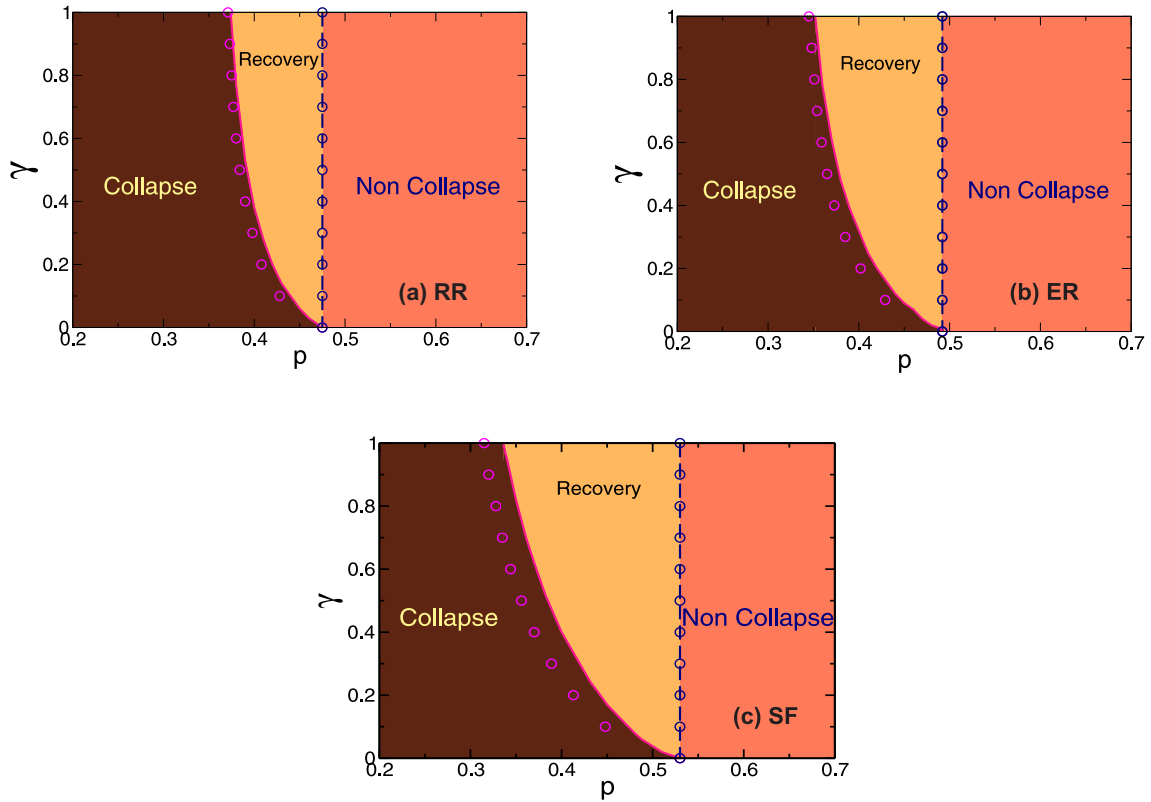}
\vspace{1.5cm}
\caption{(Color online). Phase diagrams in the $\gamma-p$ plane
  obtained from the theoretical approach and simulations for (a) RR
  networks with $z=5$, (b) ER networks with $\langle k \rangle=5$ and
  (c) SF networks $\lambda=3$ and $\langle k \rangle \approx
  5.11$. The symbols correspond to the simulations and the lines to
  the theory. The pink and magenta curves represent the values of
  $\gamma_c$ as a function of $p$. The blue lines and symbols
  represent the values of $p_c$ for $\gamma=0$.}
\label{DF}
\end{figure}
 
The relative deviation values of the simulation from the theory in the
values of $p_c$ are presented in Table~\ref{Table.1} of {\it
  Supplementary Information: Deviations of the simulated threshold
  from the theoretical }.
\begin{figure}[ht]
\centering
\includegraphics[scale=0.8]{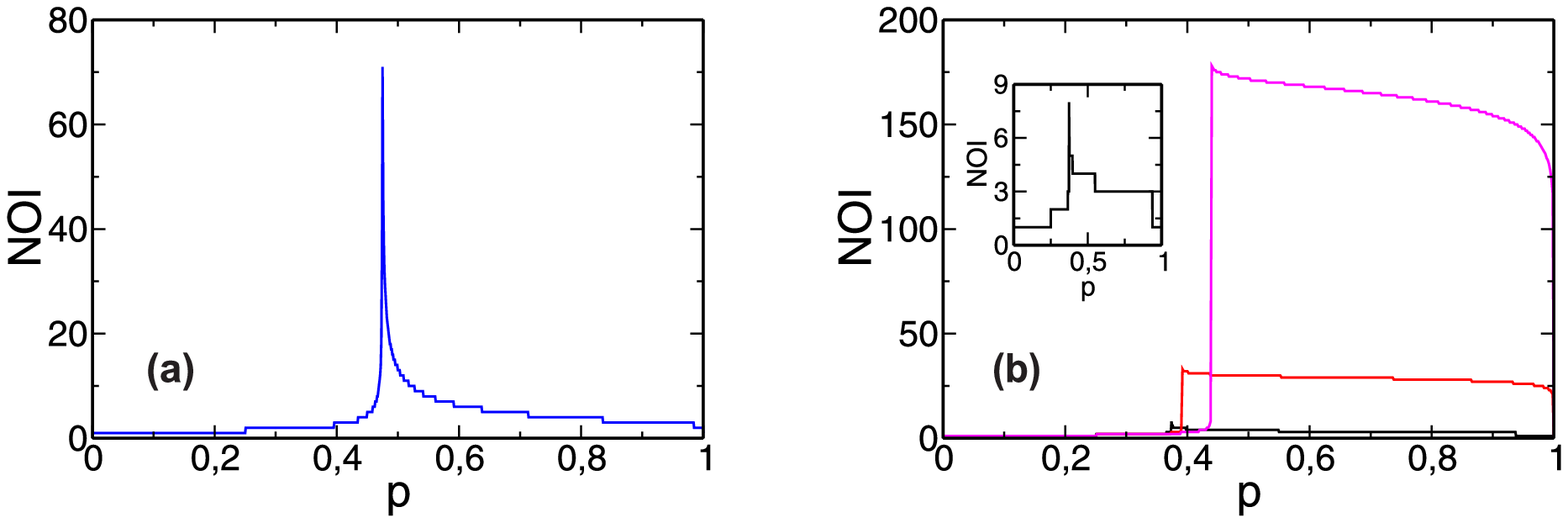}
  \caption{(Color online). Theoretical Number of Iteration Steps (NOI)
    as a function of $p$ for (a) $\gamma=0$ (blue) and (b) $\gamma=1$
    (black), $\gamma=0.5$ (red), $\gamma=0.1$ (magenta) for RR
    networks with $z=5$. The inset shows only the case $\gamma=1$ for
    a better visualization.  \label{NOI}}
\end{figure}
From the dependence of the order parameter on $\gamma$ another useful
measure can be obtained. If a system suffers a random initial removal
of a fraction of $(1-p)$ of nodes, what is the minimum value of
$\gamma$ (probability of recovery) that prevents the collapse of the
system? We denote this critical value $\gamma_c$. In Fig.~\ref{DF} we
show the phase diagrams in the $\gamma-p$ plane obtained from the
simulations and theory for RR-RR with degree $z=5$, ER-ER with
$\langle k \rangle=5$, and SF networks with $\lambda=3$ and minimum
degree $3$, which corresponds to $\langle k \rangle \approx 5.11$.
The symbols correspond to the simulations and the lines to the
theory. We can see that the relation between $p$ and $\gamma_c$ is
approximately the same for both curves, except for a small deviation
in their left sides. Hence our analysis of the phase diagram is only
based on the theory results. We can see that there are three
well-defined regions, delimited by the solid curve, which represents
the values of $\gamma_c$ for each $p$, and by the dashed line, which
indicates the value of $p_c$ for $\gamma=0$. The region located to the
right of the dashed line is the non-collapsed region, since the system
does not crash at any of these values of $p$. Note that the simulation
points coincide exactly with this curve, as the theory has an
excellent agreement with the simulations for $\gamma=0$
\cite{Bul_01}. To the left of the dashed line and up to the solid line
is the recovery phase in which there is always a minimum value of
$\gamma_c$ that prevents the collapse of the system. This region
depends on $\langle k \rangle$ and shifts to the left (lower $p$) when
the mean connectivity increases (see Fig.~\ref{SI.f} in the {\it
  Supplementary Information: Phase Diagrams}), which means that the
restoring process is needed more for lower values of $\langle k
\rangle$. Finally, to the left of the solid (pink) curve is the
collapsed phase in which the recovery process cannot prevent the
complete breakdown of the system.

\subsubsection*{Number of Iteration Steps and Dynamics}\label{SI.n}

An accurate approach that can also be used in structured networks---such
as networks with communities, degree correlation, and clustering---is to
extract the values of the critical threshold $p_c$ for each $\gamma$
from the number of iterations steps (NOI) in the cascading process,
which exhibits a maximum at $p_c$ \cite{roni_PNAS_2013}. The NOI is the
number of iterative cascade steps required for the system to reach the
steady state. It is known that in a conventional cascade of failures
without any process of recovery applied the NOI presents a very sharp
peak at the critical threshold. This means that the system requires a
long period of time to reach the steady state when $p$ is close to
$p_c$, but when we move away from $p_c$ the system reaches the steady
state in a few steps.  In Fig.~\ref{NOI} we show the theoretical values
of the NOI for RR-RR networks for $\gamma=0$, $\gamma=0.1$,
$\gamma=0.5$, and $\gamma=1$. We show only the results for RR coupled
networks because for ER and SF networks they are qualitatively the same.
\begin{figure}[h]
  \begin{center}
    \vspace{1cm}
    \includegraphics[scale=0.8]{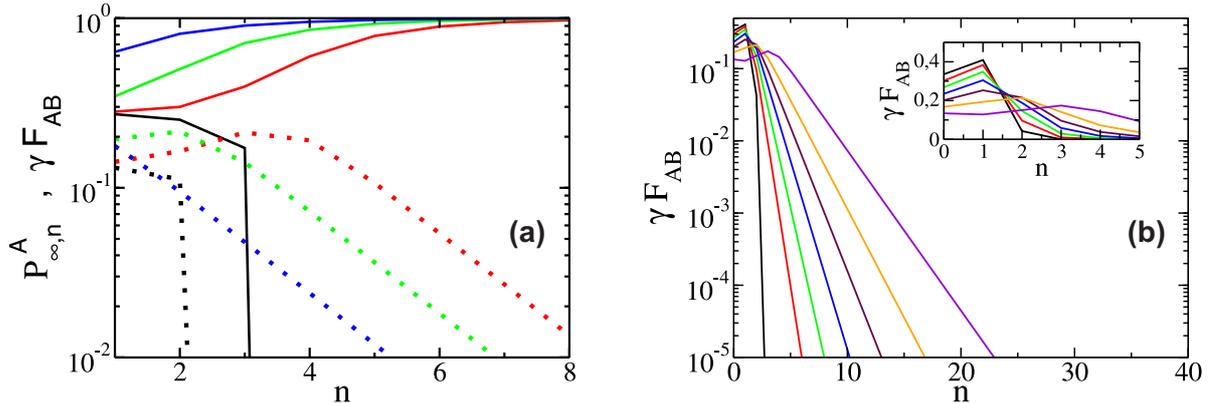}
  \end{center}
  \caption{(Color online) Log-Linear plot of (a) fraction of nodes in
    the GC of $A$, $P^A_{\infty,n}$ (solid lines) and the fraction of
    repaired nodes in the mutual boundary $\gamma \; F_{AB}$ (dot lines)
    as a function of $n$ for $p=0.391$ (black), $p=0.392$ (red),
    $p=0.4$ (green) and $p=0.48$ (blue). The black full and dotted
    lines denote the value of $p$ in the Collapse region. In red and
    green are the regions of Collapse-Recovery curves and in blue the
    Recovery region (b) fraction of repaired nodes in the mutual
    boundary $\gamma \; F_{AB}$ as a function of the iteration step $n$
    in the recovery region, with $p=0.4$ from left to right $\gamma=1$
    to $\gamma=0.4$ in intervals of $0.1$. The inset shows the maximum
    of $\gamma \;F_{AB}$ located at the first steps of the
    process. The curves where obtained for RR networks with $z=5$,
    from the theoretical approach.\label{pinfvsn}}
\end{figure}

Note that in Fig.~\ref{NOI} (a) the NOI is clearly localized in the
critical value and has a sharp peak when no strategy of recovery is
applied. From Fig.~\ref{NOI} (b) we can make two observations, (i) that
the number of steps increases as $\gamma$ decreases, and (ii) that the
NOI does not present a sharp peak and has a flattened plateau form
above $p_c$. The first observation means that as the fraction of
repaired nodes becomes larger the system requires fewer steps to reach
the steady state at the critical point, and the second indicates that
the required time for fully restoring the system is not strongly
affected by the initial failure $p$.

To better understand this behavior we show in Fig.~\ref{pinfvsn} (a)
in a log-linear scale for better visualization the temporal evolution
of the order parameter and $\gamma \; F_{AB}$ in network $A$ for RR
networks with $z=5$ and $\gamma=0.5$ for different values of $p$,
obtained from our theoretical approach. The red and blue lines
separate the three regions of the phase diagram of Fig.~\ref{DF}, the
collapsed, recovered and non-collapsed regions. Below the threshold in
the collapsed phase the system becomes dysfunctional in a few steps,
but just above the threshold there is a competition between the
recovery process and the cascade of failures and thus the number of
iteration steps greatly increases. Although the system is less damaged
in the non-collapsed region than in the recovered region the amount of
time the system needs to reach the steady state is approximately the
same. This can be explained from the temporal behavior of the mutual
restored boundary $\gamma \;F_{AB}$ shown in dashed lines in
Fig~\ref{pinfvsn} (a). As the number of steps increases and the system
approaches full restoration, the mutual boundary exhibits a peak after
which it decays exponentially. This shows that at each step of the
cascade the number of nodes repaired becomes smaller. Thus because in
our model the complete recovery of the system of networks requires
that all single nodes be reactivated, the process of recovery always
takes longer than in the collapsed region since it takes a long time
to repair the few remaining non-functional nodes.  This can be easily
seen from Fig~\ref{pinfvsn} (b) where we show a Log-Linear plot of $
\gamma \;F_{AB}$ as a function of $n$ for the same RR network as in
Fig~\ref{pinfvsn} (a) in the recovered region for different values of
$\gamma$ and $p=0.4$. Note that the fraction of recovered nodes in the
boundaries reaches a maximum in few steps, which shifts to the right
as $\gamma$ decreases. At the maximum the fraction of nodes in the GC
is almost fully recovered as shown in blue in Fig.~\ref{pinfvsn}
(a). After the maximum it decays exponentially in a characteristic
time that increases as $\gamma$ decreases, and as a consequence the
dynamic of the system takes longer to fully recover.

\section*{Discussion}

We note that the dynamics of cascading failures when $\gamma>0$ differ
greatly from when $\gamma=0$ (see Fig.~\ref{NOI} in {\it Section:
  Number of Iteration Steps and Dynamics}). The main difference is
that when $p>p_c$ and $\gamma=0$ \cite{roni_PNAS_2013} the number of
iteration steps (NOI) needed to reach the steady state decays sharply,
but when $\gamma>0$ it remains high. The reason for this difference is
that the NOI only counts cascading failure steps when $\gamma=0$, but
when $\gamma>0$ it also counts the steps of recovery to a fully
functional system. The recovered region is characterized by a dynamic
that is slower than in systems that undergo cascade of failures
without recovering.

In summary, we have proposed and studied a recovery strategy to mitigate
the breakdown of a system composed of two interdependent networks in the
presence of cascading failures. The strategy consists of repairing with
a probability $\gamma$ every node that belongs to the mutual boundary of
each GC. Our strategy yields the minimal probability, $\gamma_c$, at
which one can repair the components and prevent system collapse. We have
solved the problem theoretically using random node percolation theory
and have obtained a good agreement with the simulation results with
small deviations close to the critical point. We believe that our model
is an important contribution in developing a usable strategy for
repairing damaged infrastructure systems, and that it also suggests
future directions of research focused on recovery processes.

\section*{Methods}
\label{s.method}

For the simulations in RR-RR and ER-ER networks we use a system size of
$N=10^6$, for SF-SF $N= 5 \;10^6$ was used, and for the construction of
the networks we use the Molloy-Reed Algorithm \cite{Mol_01} averaged
over $1000$ network realizations of the process. In the simulations,
when $p$ is close to the critical value $p_c$, which depends on
$\gamma$, the network collapses in some network configurations and is
restored completely in others. For a fixed value of $p$, we consider the
system fully recovered if the network is restored in more than $50\%$ of
the realizations and collapsed if it is restored in less than 50\%. This
statement is supported by our finding that this is a finite size effect,
i.e., in the limit of infinite network size the system either collapses
or is repaired completely for a given value of $p$.

To accurately evaluate the values of $p_c$ as a function of $\gamma$
using simulations, we compute the value of $p$ at the peak of the number
of iteration steps (NOI) needed to reach the steady state
\cite{roni_PNAS_2013} (see {\it Section: Number of Iteration Steps and
  Dynamics}).

As the process of measuring the peak in the NOI requires heavy data
analysis, we compute the theoretical values of $\gamma_c$ in the phase
diagram as follows. For a fixed value of $p$ the theoretical process
is evaluated for varying values of $\gamma$. When the GCs drop to
zero, we record the $\gamma$ value to be $\gamma_c=\gamma_c(p)$. We
find numerically for all studied cases that these values coincide with
the ones obtained from the peak of the NOI.

\section{\large{Supplementary Information}}\label{S.I}
\subsection*{Theory}\label{SI.t}

In this section we explain in detail how we generalize our process
to include recovery. At the initial stage, $n=0$, the fraction of
nodes in the GC of network $A$ is given by
\[P^A_{\infty,0} = g_A[p_{0}^{A}] = g_A[p] \label{pinfaS}\ .\]

The failure next propagates to network $B$ in which the
fraction of remaining nodes is
\[p_0^B = g_A[p]\ ,\]
and hence
\[P^B_{\infty,0} = g_B[p_{0}^{B}] = g_B[g_A[p]]\ .\]

After the initial failure moves from network $A$ to $B$, the
process of recovery begins. The nodes that are repaired are those that
belong to the mutual boundary of both GCs. We calculate the
fraction of nodes that are in the border of each GC to be
\begin{eqnarray}\label{yyy01zs}
F_{0}^{A}&=&(1-p)\;(1-G^A_0 [1- f^A_{\infty,0}])\ ,\nonumber\\
F_{0}^{B}&=&(1-g_B[g_A[p]])\;(1-G^B_0 [1- f^B_{\infty,0}])\ ,
\end{eqnarray}
and the mutual boundary, given by
\[F_{0}^{AB}=F_{0}^{B}\frac{F_{0}^{A}}{1-g_A[p]}\ ,\]
where $g_A[p]$ is the relative size of the GC in network $A$ after the
cascading failure, and $F_{0}^{A}/(1-g_A[p])$ is the conditional
probability that a node belongs to the boundary of the GC of network
$A$ ,given that it is interconnected though an interdependent link with
a node that belongs to the boundary of the GC of network $B$.

We next compute the new fraction of nodes that belong to the GC
of each network
\begin{eqnarray}
\overbar{P^A_{\infty,0}} &=& g_A[p] + \gamma F_{0}^{AB}\ ,\nonumber\\
\overbar{P^B_{\infty,0}} &=& g_B[g_A[p]] + \gamma F_{0}^{AB}\ ,
\end{eqnarray}
and the fraction of functional nodes in each network after the
recovery process by solving
\begin{eqnarray}
g_A[q_{0}^{A}]&=&\overbar{P^A_{\infty,0}}\ ,\nonumber\\
g_B[q_{0}^{B}]&=&\overbar{P^B_{\infty,0}}\ .
\end{eqnarray}

We next compute the fraction of nodes remaining in network $A$ in
the next step of the cascade as in Ref.~\cite{Bul_01}
\[p^A_1 = p^A_0 \frac{g_B[q^B_0]}{g_A[q^A_0]}\ .\]
Hence the GC of $A$ at $n=1$ is given by
\[P^A_{\infty,1} = g_A[p^A_1]\ ,\]
and then the fraction of remaining nodes in $B$ is
\[p^B_1 = p^B_0 \frac{g_A[p^B_1]}{g_B[q^B_0]}\ ,\]
and the fraction of nodes in its GC can we written as
\[P^B_{\infty,1} = g_B[p^B_1]\ .\]
Then the recovery process is applied again.

\subsection*{Analytical solutions for the fraction of nodes in the GC's}

In this section we show that in the steady state when there are no
isolated nodes before the initial failure the only possible values of
the order parameter are $0$ or $1$, below and above the threshold
without intermediate states.

Note that using Eqs.~(1)--(7) in the main text we can write the temporal
evolution of the order parameters as

\vspace{1cm}

\[P^A_{\infty,n}=\overbar{P^B_{\infty,n-1}}\frac{(1-G_0^A(1-f^A_{\infty,n}))}{(1-G_0^A(1-\overbar{f^A_{\infty,n-1}})}\;,\]

\[P^B_\infty(n)=\overbar{P^B_{\infty,n-1}}\frac{(1-G_0^A(1-f^A_{\infty,n}))(1-G_0^B(1-f^B_{\infty,n}))}{(1-G_0^A(1-\overbar{f^A_{\infty,n-1}}))(1-G_0^A(1-\overbar{f^B_{\infty,n-1}}))}\;,\]

\vspace{1cm}

where $f^\alpha_{\infty,n}$ and $\overbar{f^\alpha_{\infty,n}}$ with
$\alpha=A,B$ satisfy the trascendental equations

\[f^A_{\infty,n}=\overbar{P^B_{\infty,n-1}}\frac{(1-G_1^A(1-f^A_{\infty,n}))}{(1-G_0^A(1-\overbar{f^A_{\infty,n-1}}))}\;,\]

\[f^B_{\infty,n}=\overbar{P^B_{\infty,n-1}}\frac{(1-G_0^A(1-f^A_{\infty,n}))(1-G_1^B(1-f^B_{\infty,n}))}{(1-G_0^A(1-\overbar{f^A_{\infty,n-1}}))(1-G_0^B(1-\overbar{f^B_{\infty,n-1}}))}\;,\]

\[\overbar{f^A_{\infty,n}}=\overbar{P^A_{\infty,n}} \frac{(1-G_1^A(1-\overbar{f^A_{\infty},n}))}{(1-G_0^A(1-\overbar{f^A_{\infty,n}}))}\;,\]

\[\overbar{f^B_{\infty,n}}=\overbar{P^B_{\infty,n}} \frac{(1-G_1^B(1-\overbar{f^B_{\infty,n}}))}{(1-G_0^B(1-\overbar{f^B_{\infty,n}}))}\;.\]

\vspace{1cm}

and $\overbar{P^\alpha_{\infty,n}}=P^\alpha_{\infty,n}+FAB_n$, with
$\alpha=A,B$, where $FAB_n$ is the shared boundary.

In the steady state at $n=n_s$,
$P^A_{\infty,n_s}=P^B_{\infty,n_s}$. It is straightforward to show that
$P^A_{\infty,n_s}=P^B_{\infty,n_s}=0$ is a solution of the previous
system of equations. For $P^B_{\infty,n_s}>0$ after some algebra it
can be shown that

\[\overbar{f^A_{\infty,n_s-1}}=f^A_{\infty,n_s}\;,\]

\[\overbar{f^B_{\infty,n_s-1}}=f^B_{\infty,n_s}\;.\]


Using these equalities we find

\[P^B_{\infty,n_s}=\overbar{P^B_{\infty,n_s-1}}\;.\]

\vspace{1cm}

On the other hand, it is clear that at the steady state
$P^A_{\infty,n_s}=P^A_{\infty,n_{s+1}}$. Using this relation and the
previous results we deduce

\vspace{1cm}

\[\overbar{P^B_{\infty,n_s}}=P^B_{\infty,n_s}\;.\]

\vspace{0.5cm}

Recalling that
$\overbar{P^B_{\infty,n_s}}=P^B_{\infty,n_s}+\gamma~FAB_{n_s}$, hence
we have

\[\gamma~FAB_{n_s}=0\;.\]

Thus for $\gamma>0$ the shared boundary in the steady state must be
zero. Note that the condition $FAB_{n_s}=0$ is trivially satisfied
when $P^\alpha_{\infty,n_s}=0$. Moreover, it can be shown using
L'H\^opital's rule that the condition is also fulfilled when
$P^\alpha_{\infty,n_s}=P^\alpha_{\infty,-1}$, where
$P^\alpha_{\infty,-1}$ is the original fraction of nodes in each GC
before the initial failure. On one hand, if each initial GC equals
the whole network then $P^\alpha_{\infty,n_s}=1$. On the other hand,
if before the initial random failure the probability of existence of isolated nodes
is not equal to zero ($P(k=0)\gtrapprox0$), such as in ER networks, then the
initial fraction of nodes in each GC is not the entire newtork and
thus $P^\alpha_{\infty,n_s}\lessapprox 1$.

\subsection*{Deviations of the simulated threshold from the theoretical}

The theoretical results adjust well to the simulation
results, except for small deviations when $\gamma>0$. Using the phase
diagrams in Fig.~4 in the main text, we compute these deviations as
relative errors between the theoretical and simulated values. The
relative error is defined as

\[ \epsilon_r = 1 - \frac{p_c^s}{p_c^t} ,\]

where $p_c^t$ and $p_c^s$ are the critical values obtained from theory
and simulations, respectively. Note that $p_c^s \leq p_c^t$ as explained
in the {\it Theoretical Approach\/} section in the main text. In
Table~\ref{Table.1} the relative deviations are listed for several
values of $\gamma$ and for the three types of network.

\begin{table}[h]
\begin{center}
\begin{tabular}{|c|c|c|c|}\hline

$\gamma$ &RR-RR & ER-ER & SF-SF\\ \hline\hline
   0.0 & 0.0 &0.0 &0.0 \\ \hline
   0.1 & 0.027& 0.036&0.053\\\hline
   0.2&  0.028& 0.042&0.066\\\hline
   0.3	& 0.017&0.030 &0.074\\\hline
   0.4	& 0.020 &0.045 &0.075\\\hline
   0.5	& 0.020 &0.032&0.075\\\hline
   0.6	& 0.023 &0.030&0.073\\\hline
   0.7	& 0.028 &0.027&0.072\\\hline
   0.8	& 0.016 &0.025  &0.063\\\hline
   0.9	& 0.011 &0.022&0.0067\\\hline
   1.0	& 0.008 &0.020  &0.062\\\hline\hline
\end{tabular}
\caption{Relative deviation $\epsilon_r$ of the critical theshold
  $p_c$ for different values of $\gamma$ for a system composed of two
  RR networks (second column) with $z=5$, two ER with $\langle k
  \rangle =5$ (third column) and two SF networks with $\langle k
  \rangle \approx 5.11$ (fourth column).}
\label{Table.1}
\end{center}
\end{table}

Note that the deviations do not exceed $3\%$ for RR, $5\%$ for ER, and
$8\%$ for SF. The numerical simulations give results that are very
similar to those from theory, and we now explore the interesting
features derived primarily from theory.

\subsection*{{Excess Degree of the Boundary}}\label{SI.b}

We next explain why the nodes on the boundary of the GC have higher
degrees than dysfunctional nodes that are not on the boundary. For
simplicity we will drop the network indices A and B in the main
magnitudes. The boundary of the GC is the set of nodes that have at
least one connection with the GC. The equation that represents the
relative fraction of nodes that belongs to the GC is

\begin{equation}\label{Eq.pinfs}
 P_\infty = \sum_{k=k_{min}}^{k_{max}} \tilde{p} P(k)
 (1-(1-f_\infty)^k)\;,
\end{equation}
where $f_\infty$ is the root of the self-consistent Eq.~(1) in the main
text, and $\tilde{p}$ is the fraction of remaining nodes before
repairing process is initiated.

We can rearrange the coefficients of Eq.~(\ref{Eq.pinfs}) as $P(k)
(\tilde{p}-\tilde{p}(1-f_\infty)^k)$, where $\tilde{p}$ is the fraction
of remaining nodes, and $\tilde{p}(1-f_\infty)^k $ is the probability
that a non-failed node does not belong to the GC. Since for $\tilde{p} <
1$, $f_\infty < 1$, the probability that a node belongs to a finite
cluster after failure decreases with $k$. Hence it is more likely for a
node to be part of the GC if its connectivity is fairly high.  The
fraction of nodes that belongs to the boundary is obtained by simply
replacing in Eq.~(\ref{Eq.pinfs}) $\tilde{p}$ with $1-\tilde{p}$

\[F = \sum_{k=k_{min}}^{k_{max}} (1-\tilde{p}) P(k) (1-(1-f_\infty)^k)\ . \]
Rearranging the coefficients we have
$P(k)((1-\tilde{p})-(1-\tilde{p})(1-f_\infty)^k$ where
$P(k)((1-\tilde{p})$ is the probability that a node has failed and
$(1-\tilde{p})(1-f_\infty)^k$ holds for the probability that a node has
failed and is not connected to the GC. As explained above, this last
term decreases as $k$ increases. Hence the probability that a node
belongs to the boundary of the GC increases with its degree $k$.

\subsection*{{Phase Diagrams}}\label{SI.f}
\begin{figure}[h]
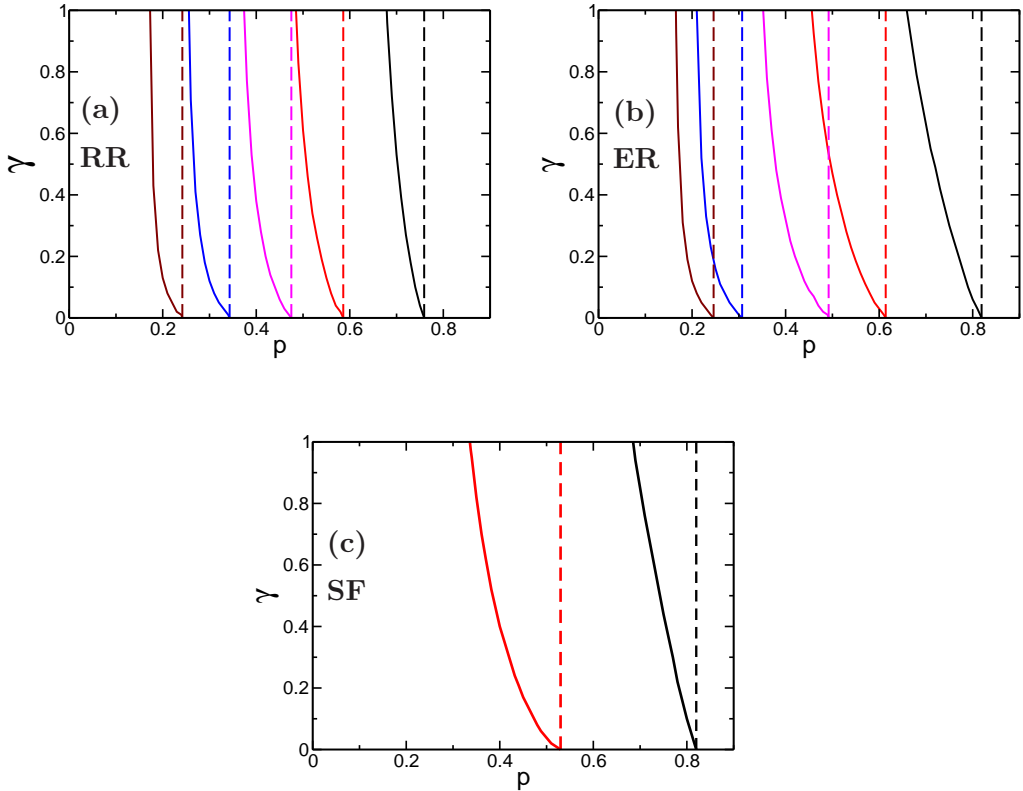

\begin{overpic}[scale=0.27]{Fig4a.eps}
    \put(15,50){{\bf{(a)}}}
\put(15,40){{\bf{RR}}}
  \end{overpic}\hspace{0.5cm}
  \begin{overpic}[scale=0.27]{Fig4b.eps}
    \put(15,50){\bf{(b)}}
\put(15,40){{\bf{ER}}}
  \end{overpic}\vspace{1cm}
  \begin{overpic}[scale=0.27]{Fig4c.eps}
    \put(15,50){{\bf{(c)}}}
\put(15,40){{\bf{SF}}}
  \end{overpic}\hspace{0.5cm}
  \caption{(Color online). Phase diagram in the plane $\gamma-p$ for
    (a) RR networks with $z=3$ (black), $z=4$ (red), $z=5$ (magenta),
    $z=7$ (blue) and $z=10$ (brown) (b) ER networks with $\langle k
    \rangle=3$ (black), $\langle k \rangle=4$ (red), $\langle k
    \rangle=5$ (magenta), $\langle k \rangle=7$ (blue) and $\langle k
    \rangle=10$ (brown). For SF networks we used $\lambda=3$ and
    minimum degrees $2$ (black) and $3$ (red). The recovery regions
    are enclosed by their respective curves. The curves were
    constructed from the theoretical values.\label{pc_gamma2}}
\end{figure}

In Fig.~\ref{pc_gamma2} we show the phase diagrams in the $\gamma-p$
plane obtained from theory for different values of $\langle k \rangle$
for RR, ER and SF networks. As indicated in the main text, the recovery
regions determined by the critical values of $\gamma_c$ for each $p$
(solid line) and by the value of $p_c$ for $\gamma=0$ (dashed line)
shift to the left (lower $p$) when the mean connectivity increases,
indicating that the restoring process is more essential when the
$\langle k \rangle$ values are lower.

Note that the recovery regions for the SF-SF networks are the broadest
for the same value of $\langle k \rangle$, i.e., of the three types of
network they have the widest range of $p$ values in which this restoring
strategy is effective. On the other hand, the RR-RR networks have the
most narrow recovery region. This difference is because the SF-SF
networks have the largest degree dispersion and in the RR-RR networks it
is null. This corroborates that large heterogeneity implies high
resilience

\section*{\it Acknowledgments}

HES thanks the NSF MMI 1125290, ONR Grant N00014-14-1-0738,
and DTRA Grant HDTRA1-14-1-0017 for financial support. SH thanks the
Defense Threat Reduction Agency (DTRA), the Office of Naval Research
(ONR) (N62909-14-1-N019), the United States-Israel Binational Science
Foundation, the LINC (Grant No. 289447) and the Multiplex (Grant
No. 317532) European projects, the Italy-Israel binational support and
the Israel Science Foundation for support. MAD, CEL and LAB wish to
thank to UNMdP, FONCyT and CONICET (Pict 0429/2013, Pict 1407/2014 and
PIP 00443/2014) for financial support. We thank L. D. Valdez for useful
discussions and support throughout this research.

\section*{\it Contributions}

All authors designed the research, analyzed data, discussed results,
and contributed to writing the manuscript. MAD, CEL and LAB
implemented and performed numerical experiments and simulations. 

\section*{Additional information}

Competing financial interests: The authors declare no competing financial interests.


\end{document}